\documentclass[3p,twocolumn,times,notitlepage]{elsarticle}

\usepackage{graphicx}
\usepackage{bm}
\usepackage{braket}
\usepackage{comment}
\usepackage{hyperref}
\usepackage{gensymb}
\usepackage{subcaption}
\usepackage{float}
\usepackage{fancyvrb}
\usepackage{xurl}
\usepackage{amssymb}
\usepackage{amsmath}
\usepackage{relsize}

\newcommand{\tens}[1]{\text{\textbf{#1}}}
\newcommand{\bs}[1]{\boldsymbol{#1}}
\newcommand\CC{C\nolinebreak[4]\hspace{-.05em}\raisebox{.4ex}{\relsize{-3}{\textbf{++}}}}

\newcounter{bla}

\journal{Computer Physics Communications}

\begin{document}

\begin{frontmatter}

\title{curvedSpaceSim: A framework for simulating particles interacting along geodesics}
\author{Toler H. Webb}
\author{Daniel M. Sussman}
\ead{daniel.m.sussman@emory.edu}
\affiliation{Physics Department, Emory University.}
\date{Spring 2023}

\begin{abstract}
A large number of powerful, high-quality, and open-source simulation packages exist to efficiently perform molecular dynamics simulations, and their prevalence has greatly accelerated discoveries across a wide range of scientific domains. These packages typically simulate particles in free (Euclidean) space, with options to specify a variety of boundary conditions. While more exotic, many physical systems are constrained to and interact across curved surfaces, such as organisms moving across the landscape, colloids pinned at curved fluid-fluid interfaces, and layers of epithelial cells forming highly curved tissues. The calculation of distances and the updating of equations of motion in idealized geometries (namely, on surfaces of constant curvature) can be done analytically, but it is much more challenging to efficiently perform molecular-dynamics-like simulations on arbitrarily curved surfaces. This article discusses a simulation framework which combines tools from particle-based simulations with recent work in discrete differential geometry to model particles that interact via geodesic distances and move on an arbitrarily curved surface. We present computational cost estimates for a variety of surface complexities with and without various algorithmic specializations (e.g., restrictions to short-range interaction potentials, or  multi-threaded parallelization). Our flexible and extensible framework is set up to easily handle both equilibrium and non-equilibrium dynamics, and will enable researchers to access time- and particle-number-scales previously inaccessible.
\end{abstract}

\begin{keyword}
molecular dynamics \sep curved space
\end{keyword}

\end{frontmatter}

{\bf PROGRAM SUMMARY}\\
\begin{small}
\noindent
{\em Program Title: curvedSpaceSimulations}   \\
{\em CPC Library link to program files:} (to be added by Technical Editor) \\
{\em Developer's repository link: https://github.com/sussmanLab/curvedSpaceSim} \\
{\em Code Ocean capsule:} (to be added by Technical Editor)\\
{\em Licensing provisions(please choose one):} GPLv3 \\ 
{\em Programming language: \CC}                                   \\
{\em Nature of problem:} 
  Molecular-dynamics-like simulations of degrees of freedom evolving on a curved two-dimensional manifold according to standard equilibrium or non-equilibrium equations of motion and interacting via geodesics.\\
{\em Solution method:} We discretize both time and space,  using modern tools from discrete differential geometry to efficiently find geodesic paths and distances. MPI parallelization is implemented to access large  system sizes, and  where appropriate  (e.g., when  dealing with short-ranged  inter-particle  potentials) we implement the ability to aggressively prune data structures, greatly decreasing the computational cost of our many-particle  simulations. \\
\end{small}

\maketitle 
\section{Introduction}
While most phenomena we observe take place in flat (Euclidean)  space, efforts to understand the behavior of systems in curved space have a rich history across many disparate fields of research \cite{tarjus2011statistical,schamberger2023curvature}. These are perhaps  most famous in the context of the Thomson problem and its generalization -- i.e., finding energy minima or maximally-separated arrangements of points on curved surfaces  \cite{thomson1904xxiv,tammes1930origin,bowick2002crystalline,agarwal2020simple,agarwal2021predicting}. In addition to the physical examples which motivated these problems, some systems in flat space can also have their properties mathematically mapped to the generalized Thomson problem. For instance, finding ground state configurations of twisted bundles of cohesive fibers can be understood as a problem of packing disks on families of curved surfaces \cite{bruss2012non,bruss2013topological}. Other, more direct, applications for  physical systems on curved two-dimensional manifolds can be found in settings ranging from the adsorption of particles on porous media or other highly curved substrates \cite{pincus1984polymer,hanke1999critical}, colloids trapped at curved fluid-fluid interfaces \cite{liu2016curvature}, or the behavior of epithelial cells that form curved monolayers \cite{goodwin2019smooth,yu2021adaptive,chang2022quantifying,marin2023mapping, luciano2024mechanoresponse}. In a more indirect attempt to theoretically understand challenging problems, the curvature of space is sometimes used to tune relative levels of geometric frustration, as in explorations of disordered solids embedded in positively or negatively curved two- and three-dimensional spaces \cite{sausset2008tuning,turci2017glass}. 

Recent work -- particularly  in soft and living material systems -- has pointed towards the novel ways that curvature can affect collective dynamical and mechanical properties. For instance, theoretical and computational work has indicated the possibility for novel phenomena when considering flocking on curved surfaces \cite{activesphere2015,activepart2016,topsound2017}, which may be related to recent observations of unusual velocity waves induced by curvature on the surface of cell spheroids \cite{brandstatter2023curvature}. Coupling between growth, deposition, and curvature has been proposed as a robust route to surface patterning in pollen grains, spores, and insect cuticles \cite{lavrentovich2016first,radja2019pollen}. Experimental work has shown that cells can have the modulation of their shape \cite{luciano2021cell} and their motility \cite{pieuchot2018curvotaxis,gehrels2023curvature}  mediated by  curved surfaces, and that cells can both sense and use these curvature cues to regulate remodeling and collectively migrate \cite{maechler2019curvature,luciano2021cell,tang2022collective}.

There is clearly a need for computational tools to help study many of these  processes, but there are substantial challenges in studying the evolution of particles on curved spaces \cite{tarjus2011statistical}. Of particular note are constraints (either implicit or explicit) needed to keep the degrees of freedom on the curved manifold, and the challenge of computing interactions not according to Euclidean distances but according to the \emph{geodesic} curve separating interacting particles. One fruitful approach has been to consider instead the solution of continuum (e.g., phase field) models on curved surfaces, as in recent work on cylindrical \cite{happel2022effects} or largely  arbitrary surfaces \cite{rank2021active,hueschen2023wildebeest}. Alternatives consider particle based simulations on surfaces of constant positive or negative Gaussian curvature, where the geodesic distance calculations can be done analytically and efficiently \cite{activesphere2015,sussmancurvrigid2020, thomas2023shape}. Other approaches include constraining particles to the surfaces and using sufficiently long-range Euclidean interactions so that particles on the surface correspond to energetic ground states \cite{giomi2008elastic} (motivated by the so-called ``poppy-seed bagel theorem'' \cite{hardin2005minimal}), or simply constraining particles to the surface via projection operators while computing force vectors along Euclidean separations between particles \cite{tarjus2011statistical,schonhofer2022curvature}. All of these approaches have been productive, but have left unfilled a need to study the curved-space dynamics directly.

\begin{figure}
    \centering
    \includegraphics[width = 1.0\columnwidth]{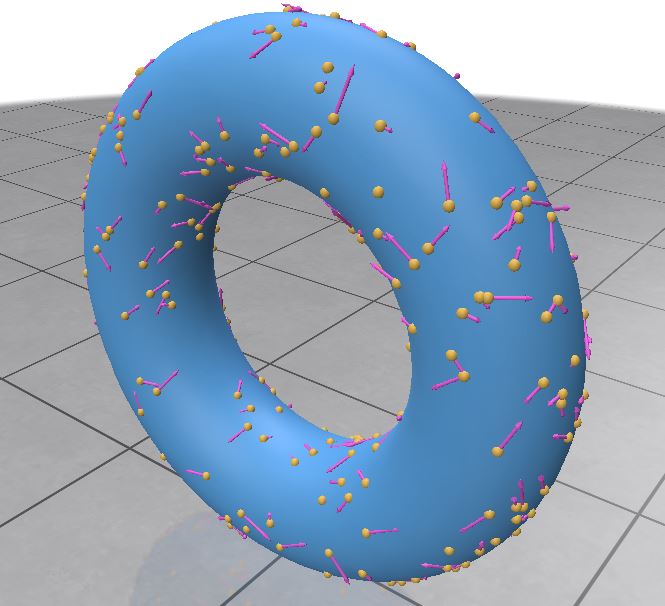}
    \caption{Snapshot from a real-time user interface demonstrating the NVE dynamics of 300 particles on the surface of a torus. In this example, particles interact according to a  soft harmonic repulsion along geodesic curves, and velocity vectors are parallel transported on geodesics as the particles move. Visualization provided via the Polyscope package \cite{polyscope}.}
    \label{fig:UIdemo}
\end{figure}

In this work we address this need by describing a flexible, extensible, open-source computational framework for efficiently performing particle-based simulations on non-self-intersecting curved surfaces. We focus for ease and clarity on closed surfaces with otherwise arbitrary geometry and topology, but note that extending our framework to include, e.g., open surfaces or surfaces with periodic boundary conditions is in principle straightforward. While at this stage we are focused on command-line interfaces to the underlying code, we have integrated some of the framework with a graphical user interface, as shown in Fig.~\ref{fig:UIdemo}.

The rest of this article is organized as follows. In Sec.~\ref{sec:background} we first review some of the core ideas of performing molecular-dynamics-like simulations, and in particular focus on a division of the problem into classes that perform the usual work of evaluating forces and integrating various equations of motion \cite{frenkel2023understanding} and those that handle metric spaces and the core  ``shift'' and ``displacement'' functions \cite{schoenholz2021jax} that differ from one curved surface to another. We next review the essential concepts and tools from discrete differential geometry that we use to construct these more specialized classes. In Sec.~\ref{sec:methods} we discuss some of our algorithmic implementations, and in Sec.~\ref{sec:benchmarks} we show performance data, focusing on the scaling of computational cost with increasing particle number and on surfaces of increasing complexity (along with the strong scaling of parallelizing these calculations across multiple threads). We demonstrate a simple application of our simulation framework (related to the generalized Thomson problem on the torus) in Sec.~\ref{sec:example}, and close with a brief discussion of future directions in Sec.~\ref{sec:conclusion}.

\section{Background}\label{sec:background}

\subsection{Particle-based simulations}
Molecular dynamics and related methods discretize time to solve for the evolution of large numbers of degrees of freedom interacting according to pairwise or many-body forces and updating according to one of any number of equations of motion \cite{frenkel2023understanding}. Common choices include integrating Newton's law of motion, 
\begin{equation}\label{eq:basicupdate}
\bs{f}_i = m_i \bs{a}_i,
\end{equation}
where $\bs{f}_i$ is the force on particle $i$, $m_i$ is its mass, and $\bs{a}_i$ its acceleration. One can discretize time via, e.g., a velocity  Verlet update scheme to simulate particles in the NVE ensemble, using any number of thermostats and barostats to specify alternate statistical ensembles. Particle motion can also be coupled to a strongly dissipative background, as in the case of discrete Brownian dynamics (a limiting form of Langevin dynamics),
\begin{equation}
\lambda \Delta  r_{i,\alpha} = (\Delta t) f_{i,\alpha} + \sqrt{2 k_B T \lambda \Delta t }\eta_i,
\end{equation}
where $\alpha$ denotes a Cartesian component, $\lambda$  is a frictional term and the noise $\eta_i$ is delta-correlated white noise with $\langle\eta\rangle=0$ and $\langle \lvert\eta\rvert^2 \rangle = 1$. Active systems can be simulated by using any number of out-of-equilibrium  dynamics, such as self-propelled particle dynamics \cite{mishra2010fluctuations}, or via active fluctuations in mechanical stresses \cite{yamamoto2022non}. Non-physical dynamics can be imposed to measure transport coefficients by imposing artificial fluxes (e.g., of momentum) on the degrees of freedom \cite{muller1997simple}. 

This list of variations could go on, and many object-oriented simulation packages have been developed to perform one or some number of these flavors of particle-based simulations while easily swapping between different equations of motion and particle-particle interactions. At heart, all of these methods rely on a small number of algorithmic primitives: the ability to calculate forces (typically by considering gradients of a conservative potential), and the ability to update state information (updating the position or velocity of a particle, for instance). Computing forces is straightforward in principle, and in practice much of the historical algorithmic development in the community has centered on efficiently computing forces for large numbers of interacting particles -- Verlet lists, cell lists, and similar accelerating structures for short ranged potentials  \cite{howard2016efficient}, Ewald summation for long-range potentials under periodic boundary conditions \cite{toukmaji1996ewald}, and so on. Interesting variations sometimes include the addition of non-equilibrium aligning ``forces'' \cite{vicsek1995, gregoire2004onset, cavagna2015flocking} or forces representing specialized interactions found in living materials  \cite{dombrowski2004self,sussman2017cellgpu, vuijk2021chemotaxis}, but this typically induces at most a  modest increase in the difficulty of performing simulations. The second core algorithmic need -- updating state information -- is so trivial that typically little thought is given to it: particle positions and velocity updates simply amount to vector addition, potentially with some  additional bookkeeping when periodic boundary conditions are employed.

Concretely, consider a pairwise interparticle potential corresponding to a repulsive Gaussian,
\begin{equation}
    U(l_{ij}) = A e^{-\frac{l_{ij}^2}{2 \sigma^2}},
\end{equation}
where $A$ determines the stiffness of the interaction, $\sigma$ is a measure of the length scale over which the repulsive force acts, and $l_{ij}$ is the distance between particles $i$ and $j$. The force corresponding to this potential is
\begin{equation}\label{eq:f1}
    f_{ij} = -\nabla U = \frac{l_{ij}}{\sigma^2}U(l_{ij})\nabla l_{ij}. 
\end{equation}
In a typical particle-based simulation in Euclidean space, both the distance $l_{ij}$ and its gradient are easily calculable. Using $r_{i,\alpha}$ to represent the Cartesian coordinate $\alpha$ of particle $i$, one  simply  uses the appropriate Euclidean metric for length,
\begin{equation}
    l_{ij} = \sqrt{\sum_\alpha{\left(r_{j,\alpha}- r_{i,\alpha}\right)^2}},
\end{equation}
and notes that the gradient is the unit vector tangent to the line between the two particle positions: 
\begin{equation}
    \nabla_\alpha l_{ij} = \frac{(r_{j,\alpha} - r_{i,\alpha})}{\sqrt{\sum{\left(r_{j,\alpha}- r_{i,\alpha}\right)^2}}}. 
\end{equation}
From Eq.~\ref{eq:f1} we see that generalizing the calculation of a force from Euclidean to curved spaces corresponds to generalizing the calculation of \emph{distances} and \emph{gradients of distances}. We will refer to the joint task of computing these two quantities as the ``\verb+displacement+'' algorithm, and will refer to the task of updating state information (either moving a particle position along a geodesic that points in some direction, or parallel transporting a particle's velocity vector along the same path) as the ``\verb+shift+'' algorithm. 

\subsection{Discrete differential geometry}
In this software package we focus our attention on simulating particles constrained to curved surfaces and interacting along geodesics -- the so-called ``curved-line-of-force'' problem \cite{tarjus2011statistical,post1986statistical}. Here we review the basic (discrete) differential geometry employed in our simulations, and refer interested readers to Refs.~\cite{kamien2002geometry,needham2021visual} for more thorough treatments. 

In curved space, the first quantity of interest is the equivalent of the distance $l_{ij}$ between points $i$ and $j$. The challenge is to find \emph{geodesics} -- paths which are straightest and locally shortest curves between two points on the surface. If a path on the surface -- $\textbf{r}(s)$, where $s$ is an arc-length parameterization of the path -- is already known, the distance along that path between points can be expressed as an integral,
\begin{equation}\label{eq:distance}
d_{ij} = \int ds\sqrt{g_{\alpha\alpha} \left(\frac{dr_\alpha}{ds}\right)^2 + g_{\beta\beta}\left(\frac{dr_\beta}{ds}\right)^2 + 2g_{\alpha\beta}\frac{dr_\alpha}{ds}\frac{dr_\beta}{ds}}.
\end{equation}
Here  $r_\alpha(s)$ and $r_\beta(s)$ are the intrinsic coordinates of the path on the surface. The \emph{metric tensor} $\tens{g}$ contains the information needed to define distances and angles in the tangent space of the surface at every point. The components of the metric tensor are given by 
\begin{equation}
    g_{\alpha\beta} = \partial_\alpha \bs{R} \cdot \partial_\beta \bs{R},
\end{equation} 
where $R(s_\alpha,s_\beta)$ is a parameterization of the surface in Euclidean space. For example, given a surface defined by $\bs{R}(x,y) = (x,y,\sin{x})$, we have $\partial_x \bs{R} = (1,0,\cos{x})$, $\partial_y \bs{R} = (0,1,0)$, and  hence 
\begin{equation}
    \tens{g} = \left(\begin{matrix}
        1+\cos^2{x} & 0 \\ 
        0 & 1
    \end{matrix}\right). 
\end{equation}

Finding the geodesic (and also the geodesic distance) between the two points corresponds to extremizing Eq.~\ref{eq:distance} over all possible paths between the two points. This is equivalent to solving the ``geodesic equation,'' a second order ordinary differential
equation given by
\begin{equation}\label{eq:geoeq}
    \ddot{r}^\gamma + \Gamma_{\alpha\beta}^{\gamma}\dot{r}^\alpha \dot{r}^\beta = 0.
\end{equation}
Here $\Gamma$ represents the Christoffel symbols, themselves functions of the metric tensor,
\begin{equation}
\Gamma_{\alpha\beta}^{\gamma} = \frac{1}{2}g^{\gamma\zeta}\left(\frac{\partial g_{\beta\zeta}}{\partial r^\alpha} +\frac{\partial g_{\alpha\zeta}}{\partial r^\beta}-\frac{\partial g_{\alpha\beta}}{\partial r^\zeta}\right), 
\end{equation}
which help describe vector transport. In the above expressions Greek indices correspond to intrinsic surface coordinates, subscripts and superscripts correspond to covariant and contravariant quantities, and we use Einstein summation convention for repeated indices.
While possible, the direct method of solving the geodesic equation for every pair of points in the simulation, and then integrating over the found geodesic paths to find distances, is hopelessly slow -- replacing a simple Euclidean norm of the difference of two vectors with the solution to a 2nd order ODE followed by a numerical integration will not permit simulations with large numbers of particles. Evidently, the direct approach is reasonable only when the geodesic equation can be analytically solved, but this is limited to extremely specialized geometries (i.e., spheres and the hyperbolic plane). Indeed, even for the relatively  simple case of a torus the solutions to the geodesic equation are surprisingly complicated, and not all torus geodesics can be described analytically \cite{jantzen2012geodesics}. 

To make progress, then, we formulate our simulations with a discretization of both time \emph{and} space, allowing us to use the tools of discrete differential geometry \cite{crane2018discrete}. We replace the smooth curved surface of interest with a triangulated version with $V$ total vertices. Just as discretization in time leads to deviations from the true solution to the equations of motion -- where the order in the error depends on the specific algorithm (e.g., $\mathcal{O}(\Delta t)$ for an Euler time-stepping method, or $\mathcal{O}(\Delta t^2)$ for a straightforward velocity  Verlet integrator, or...) -- this discretization of space will lead to errors in the calculation of distances and forces and displacements that are typically  of order  $\mathcal{O}(h^p)$, where $h$ is a characteristic length scale of the triangles in the mesh and $p$ is a number that will be sensitive to the precise discrete-differential-geometry algorithms used. 

A large number of techniques for computing geodesic lengths in the discrete setting have been proposed; see Ref.~\cite{crane2020survey} for a recent survey. Different techniques fall into either exact or approximate categories -- exact methods possess guarantees about finding true geodesics \emph{on the approximated surface} whereas approximate methods do not  but are typically  faster. This corresponds to effectively different values $p$ for the order of error made by the discretization of the surface in various calculations. For the problem of finding geodesic distances, e.g., Ref.~\cite{sharp2019vector}  suggested an $\mathcal{O}(h^1)$ scaling for the approximate ``vector heat method'' and $\mathcal{O}(h^2)$ for a competing exact method (described below).

We ultimately settled on a version of one of the exact algorithms, and in particular one which returns not only the geodesic distance between points but also the full discrete path. First proposed in Ref.~\cite{mitchell1987discrete}, the ``MMP'' algorithm  finds straightest, shortest paths between points by considering all possible local unfoldings of the mesh into a single plane and then looking for the shortest path between the two points of interest that stays within the unfolded mesh. Substantial improvements to the core algorithm -- adding priority queues for promising unfolding orders, aggressively pruning away unfoldings which cannot contain the correct path, and managing memory \cite{chen1990shortest, surazhsky2005fast} -- have made it quite performant, and a form of the latest Xin and Wang (XW) \cite{xin2009improving} version of MMP can be found in the Computational 
Geometry Algorithms Library (CGAL) \cite{cgal:eb-23b}. MMP and its variants are ``single-source-all-distance'' algorithms -- given a source point they consider all possible unfoldings and find for essentially the same amount of computational effort the distance from the source point to \emph{all vertices of the mesh}. This makes them particularly appropriate for the task of constructing neighbor lists by finding, e.g., the distances to all neighbors of particle $i$ simultaneously. We note, however, that the vector heat method mentioned above may be sufficiently faster so that using it with much finer meshes (to compensate for the $\sim\mathcal{O}(h)$ difference in discretization error) may ultimately be more efficient overall.

\section{Methods and implementation}\label{sec:methods}
As noted above,  many of  the common components in frameworks for performing particle-based simulations are quite standard \cite{frenkel2023understanding}. Our object-oriented \CC{} code  uses a standard shared-pointer  paradigm in which a governing \verb+simulation+ object connects  a  \verb+model+ containing state information (particle positions and velocities) with \verb+updater+s that implement specific equations of motion and  also with \verb+force+s by which particles interact. This basic structure (plus associated  utility classes) makes it straightforward to implement new equations of motion, new pairwise potentials, and other more exotic interactions between degrees of freedom. It also cleanly separates the standard components of a particle-based simulation from the more unusual \verb+displacement+ and \verb+shift+ algorithms we  need to implement on discretized surfaces. We define \verb+space+ classes that implement these algorithms -- either the trivial versions in Euclidean spaces or the discrete differential geometry analogs mentioned above in ``Mesh'' spaces -- and then connect a  \verb+model+ object to a \verb+space+ object its degrees of freedom live in. When particles live in a Euclidean space,  positional information is stored directly as a location in that space; when particles are on a mesh we overload the relevant data structure to contain an index for the face containing the particle and a triple of doubles indicating the position of the particle within that face in barycentric coordinates \cite{crane2018discrete}.

\subsection{Calculating distances, gradients, and forces}
The XW algorithm is a single-source-all-distance approach that requires a significant  computational cost to create a \emph{sequence tree} data structure (related to the set of possible unfoldings of the mesh relative to the source). Once created, this structure can be queried to calculate both discrete geodesic distances to any other point in the mesh and also the full geodesic path corresponding to that distance. The sequence tree has a computational complexity that is in principle $\mathcal{O}(V^2 \log V)$ to build, where $V$ is the number of vertices in the mesh. Interestingly, in practice it seems that this worst-case complexity rarely holds for most meshes \cite{xin2009improving}, as is evident from our performance analysis in the next section. 

As  mentioned above, the calculation of forces reduces to the operation of the \verb+displacement+ algorithm. Given a source particle, we identify all potentially interacting particles as ``targets,'' generate the sequence tree for the source particle, and then query the cached sequence tree to obtain the paths from the source to each of the targets. This gives the lengths $l_{ij}$ and the tangent directions $\dot{\bs{r}}_{ij}$ relative to the source point's location, from which one can (e.g.) compute the total force on $i$ via
\begin{equation}
    F_i = \sum_{j=1}^N{-\frac{dU\left(l_{ij}\right)}{d l_{ij}} \dot{\bs{r}}_{ij}}.
\end{equation}

For particles interacting via short-ranged forces, this approach contains an obvious inefficiency. The XW scales with the \emph{total} number of vertices of the discretized surface, but particles may only be interacting in a much smaller local patch of the surface. For short-range interactions, then, we perform a \emph{submeshing} operation: for a potential that vanishes beyond some distance $\sigma$, we exploit the fact that the Euclidean distance between two points is a \emph{lower bound} to the geodesic distance. For a given source particle, we consider the local patch of the surface which contains only those faces (and target particles) whose minimum Euclidean distance is within $\sigma$ of the source particle position, and then build the XW sequence tree only for these local patches. Since the local patch might contain a number of vertices $V' \ll V$, this can speed up the computation of each time step by orders of magnitude. 

\subsection{Updating state information}
The simpler of the two main algorithmic tasks, in flat space and on curved surfaces, is the \verb+shift+ update of positions and velocities of particles given a displacement vector. Force and velocity vectors are always defined in the tangent plane of the particle's current face, and we require (a) that all requested displacements are likewise in the tangent plane and (b) that at all times the particles remain on the surface. One strategy for achieving this goal is to use  a projection operator approach: for instance, a particle would be displaced by $\bs{d}$ via standard vector addition (which is, indeed, the entirety of the the \verb+shift+ algorithm in flat space), and then projected back to the nearest point on the discretized surface. While straightforward, this leads to displacements that do not themselves follow geodesics -- i.e., they are not the equivalent of a ``straight'' displacement.

\begin{figure}[th]
    \centering
    \includegraphics[width = 1.0\columnwidth]{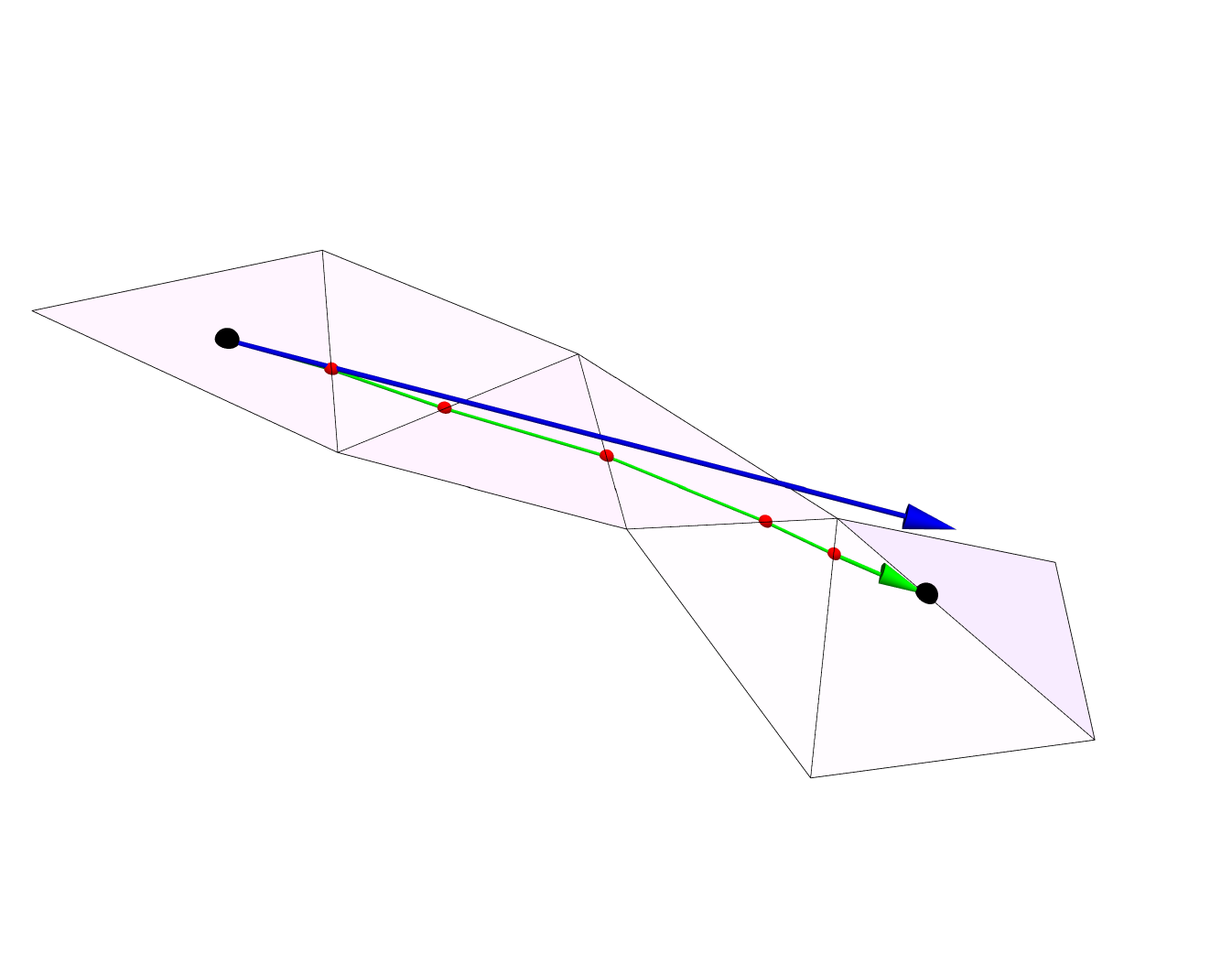}
    \caption{Operation of the \texttt{shift} algorithm: an initial position, marked by the left-most large black point, is moved by a target displacement of direction and magnitude indicated by the blue arrow. The in-surface green arrow represents the progressive rotation of the displacement vector around intersected edges in order to follow a discrete geodesic, which terminates at the right-most large right-most black point. Intersections between the path and the edges are marked by smaller red points. In-face displacements and rotations around edges continue until the total target displacement has been achieved.}
    \label{fig:wraparound_shift}
\end{figure}

Instead, as is the standard alternative, we move across connected faces along a path whose total length is the same as the magnitude of the desired initial displacement. Every time an edge is crossed, the path is rotated around the axis defined by the cross product of the face normals, as shown in Fig.~\ref{fig:wraparound_shift}. This strategy is equivalent to first defining a vector corresponding to the direction and magnitude of desired motion and then unfolding a sequence of faces so that the vector stays entirely within a connected set of faces -- i.e., of shifting the particle along  a geodesic. When crossing through a \emph{spherical vertex} --that is, a vertex
for which the sum of edge-edge angles is less than $2\pi$ -- the straightest path is no longer the shortest path on the surface, and we choose to displace particles along locally  straightest paths  \cite{polthier2006straightest}.

The main subtlety in the above algorithm is that for a given point and displacement vector, it is not \emph{a priori} clear whether or how many face crossings will be implicated. In our \verb+shift+ algorithm we implement a while loop that (a) checks for intersections -- in barycentric coordinates, to minimize finite-precision issues in these geometrical calculations -- between a  vector emanating from the  particle's current location and an edge of its current face, (b) moves the particle along that vector to either the final location (if in the face) or the intersection point, (c) if necessary, updates the face associated with the particle to the face that shares the intersected edge with the original face, and (d) rotates the vector about the intersected edge into the plane of the new face. This procedure is repeated until the total displacement has the correct magnitude (with appropriate cases handling intersections with vertices rather than edges). This loop has the benefit of easily handling parallel transport of vectors as well as shifts of particle positions: a velocity vector simply gets carried along with the position, rotating along with the rotation of the displacement vector  as edges or vertices are crossed.

\section{Performance Benchmarks}\label{sec:benchmarks}
We benchmarked our simulation architecture on a Dell OptiPlex 7080 workstation with an Intel
i9-10900K CPU clocked at 3.70 GHz. All benchmarking and example simulations were done on 
CPU using a single thread, except where noted otherwise.

\begin{figure}[t!]
    \includegraphics[width=\columnwidth]{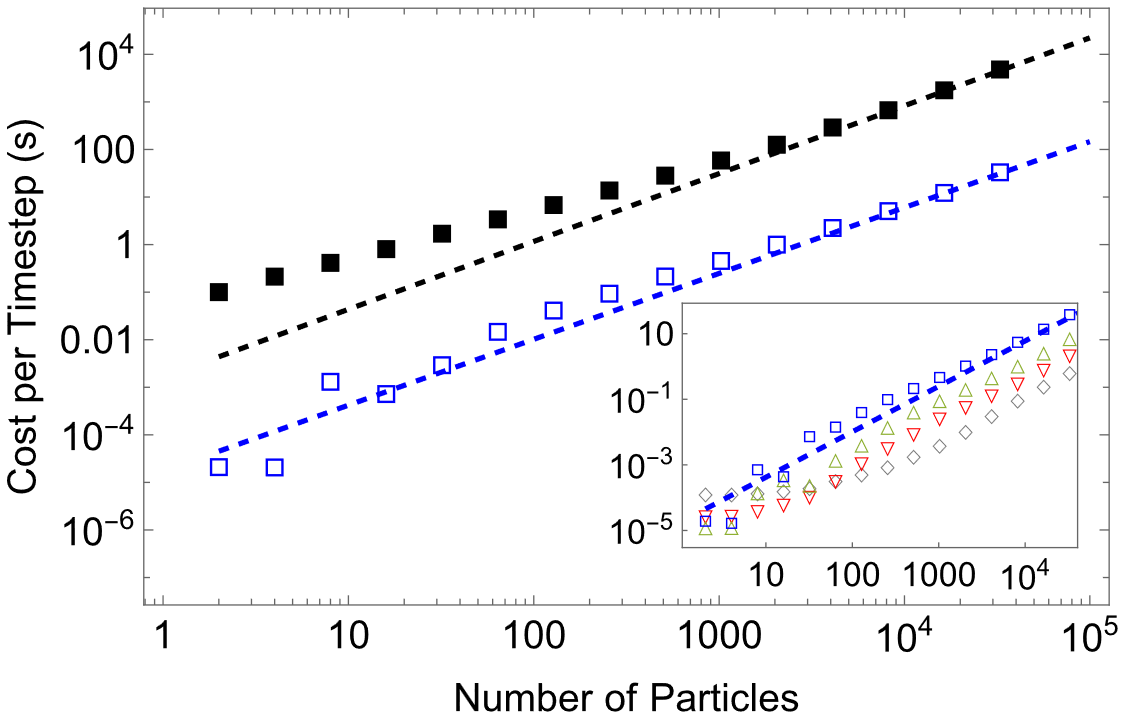}
    \includegraphics[width=\columnwidth]{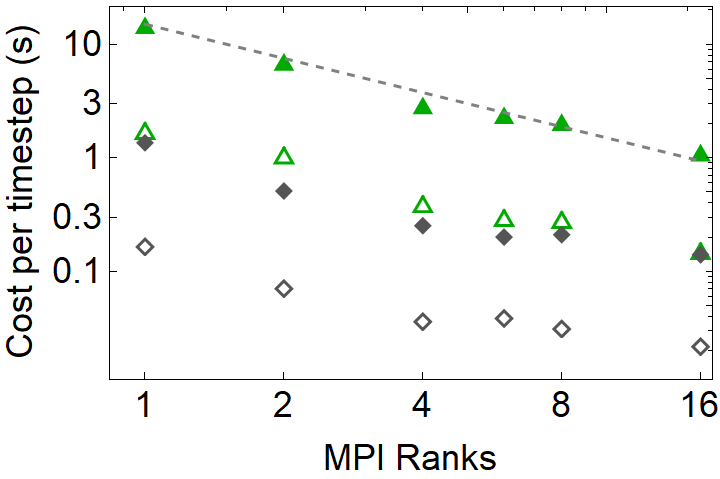}
    \caption{Simulation cost on a fixed (toroidal) mesh. (Top) The main figure shows the time per time step for a gradient descent algorithm as a function of the number of particles on the surface. The filled  black squares correspond to calculating sequence trees for the entire mesh, whereas the open blue squares correspond to using a submesh scale of $\sigma=1$. Lines reflect fits to the the asymptotic behavior of the two strategies, which both behave as power laws with scaling exponent approximately $1.4$. The inset shows the simulation cost as a function of varying $\sigma$. From top to bottom the symbols correspond to $\sigma = 2^{-n}$ for $n=0,1,2,3$, with the dashed line from the main image reproduced for convenient comparison. (Bottom) Strong scaling analysis, using a basic MPI implementation to partition the workload of updating an NVE equation of motion on the same torus across multiple processors. Green triangles and gray diamonds correspond to $\sigma=2^{-1}$ and $\sigma=2^{-3}$ as in the above top panel, and filled vs open symbols correspond to particle number $N=2^{15}$ and $N=2^{13}$.}
    \label{fig:computationalComplexityFixedMesh}
\end{figure}

We first test the performance of our algorithm on a fixed surface at fixed discretization scale (in this case, a torus with minor radius $r_1=1$ and major radius $r_2=3$, discretized into $4616$ triangular faces), as a function of the number of particles, $N$. For the main test we considered a finite-range repulsive pairwise potential whose interaction range was set to $\sigma=1$, and we computed the mean time to execute a timestep of velocity Verlet dynamics. For a standard (flat space) simulation \emph{at fixed density} one would expect $\mathcal{O}(N \log N)$ scaling; in a domain of fixed spatial extent the number of interacting particle pairs itself grows with $N$, and we would anticipate  roughly $\mathcal{O}(N^{1.5})$ scaling for particles that attempt to uniformly cover the curved surface. This is roughly consistent with the observed scaling shown in Fig.~\ref{fig:computationalComplexityFixedMesh}.

We also have implemented a basic MPI communication pattern using the openMPI framework \cite{gabriel2004open} pattern to partition work across multiple processors. The cost of computing even submeshed sequence trees is substantially higher than the cost of the mere vector addition and subtraction needed to compute distances and gradients in flat space, and we observe reasonable strong scaling behavior of our implementation at very modest system sizes. As shown in Fig.~\ref{fig:computationalComplexityFixedMesh}, we estimate that for simple meshes and relatively short-ranged potentials, assigning each processor of order $10^3$ particles almost completely hides the communication cost and results in ideal strong scaling.

\begin{figure}[hbt]
\centering
\includegraphics[width=\columnwidth]{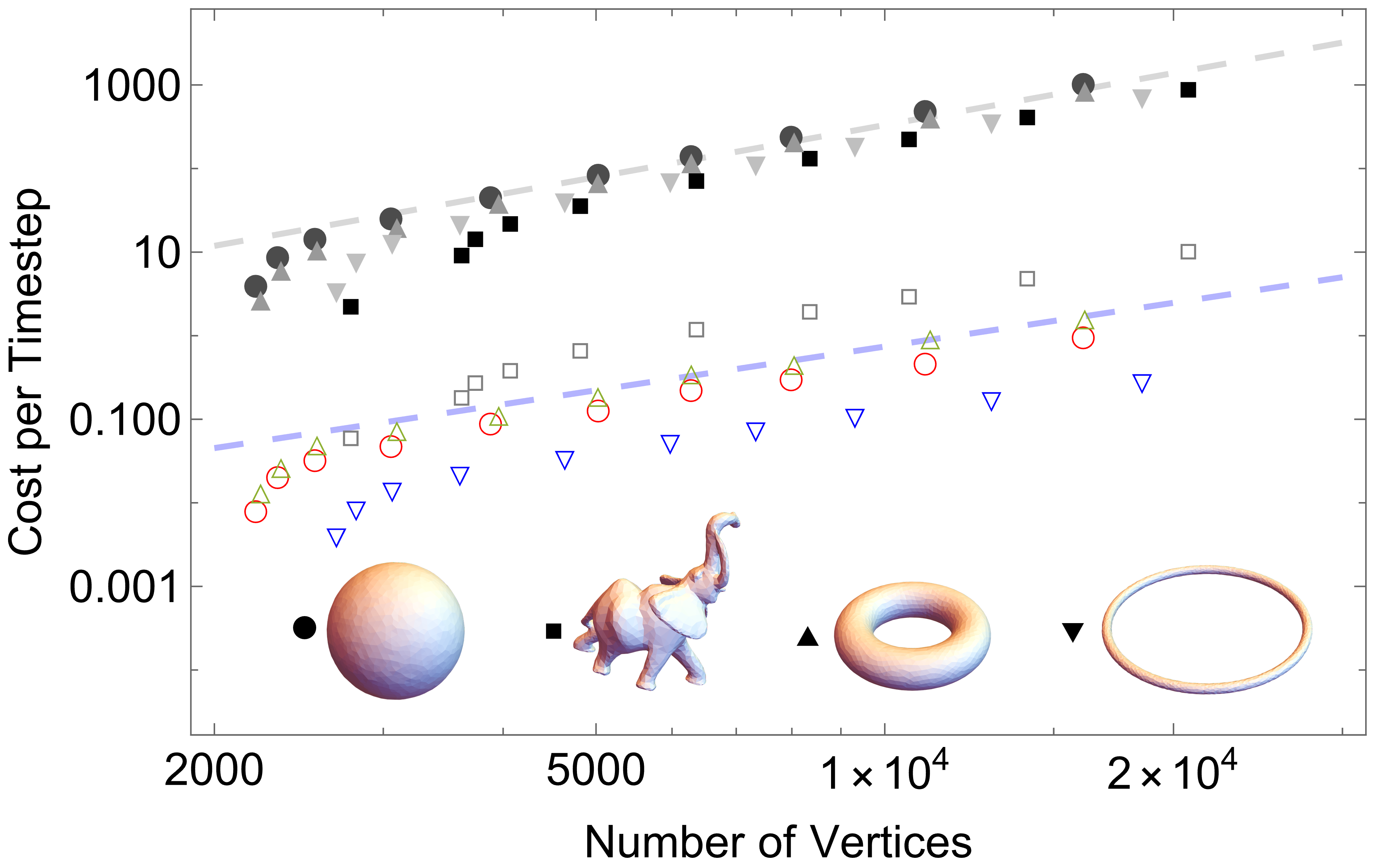}
\caption{Mean time per NVE timestep for $N=100$ particles on four surfaces upon isotropically increasing the number of vertices in each mesh. Solid symbols correspond to calculating sequence trees over the entire mesh for every particle, and open symbols correspond to using our submeshing routines for interactions of fixed range. For all surfaces, we chose an interaction range corresponding to one-tenth of the total span of the mesh in order to fairly compare across very different surface geometries.
The upper dashed line has slope $m\approx 2.07$, while the lower line has slope $m\approx 1.73$; each is an approximate fit to the the aspect-ratio-3 torus data.}
\label{fig:computationalComplexityFixedParticleNumber}
\end{figure}

The counterpart to these  analyses is to consider simulating a fixed  number of particles on meshes of increasing numbers of vertices -- this corresponds to the spatial version of choosing a time-step size that sufficiently minimizes discretization error. In addition to the $\mathcal{O}(V^2 \log V)$ worst case scaling for building the sequence tree, there are also geometric prefactors (the length scale of curvature variations, the density and placement of saddle vertices, etc) that can strongly influence the cost of these calculations. Some benchmarks have even reported non-monotonic scaling in the sequence tree generation time as a function of vertex number in some regimes \cite{cgal:klcdv-tsmsp-23b}. In Fig.~\ref{fig:computationalComplexityFixedParticleNumber}  we report the mean time to complete a timestep as a function of  $V$ for a small number of  representative surfaces: a sphere (for which all calculations  can be compared with the analytical result), tori of two different aspect ratios (3 and 20), and an elephant (in this case, represented as a geometrically interesting genus-3 surface). In all of these tests we hold the mesh span and the range of interactions fixed while 
increasing the number of triangles used to discretize the surface with an incremental triangle-based isotropic remeshing algorithm \cite{botsch2004remeshing}. We indeed find experimentally that there can be important differences when simulating particles on different meshes. Reflecting the local flattening that submeshing can accomplish we find very roughly $\mathcal{O}(V^{1.5})$ scaling for finite-range simulations compared to roughly $\mathcal{O}(V^{2})$ behavior for computing sequence trees across the entire mesh.

\section{Crystallization on the surface of a torus}\label{sec:example}
In this section, we present a simple example of using direct gradient descent methods to search for local minima of particles interacting via a purely repulsive potential on the surface of the torus. The torus is often taken as one of the simplest examples of a topologically non-trivial surface with gradients in the surface curvature \cite{bowick2009two}, and crystallization on their surface has been studied both theoretically and numerically \cite{giomi2008elastic,giomi2008defective}. We make some direct comparisons with those earlier numerical studies, which we note were implemented with forces between particles interacting according to Euclidean (rather than geodesic) distances, using a long-range Riesz potential ($U(l_{ij}) = l_{ij}^{-3}$) to ensure that in ground states particle positions were on or near the surface of the torus.

\begin{figure}[b!]
\centering
\includegraphics[width=0.75\columnwidth]{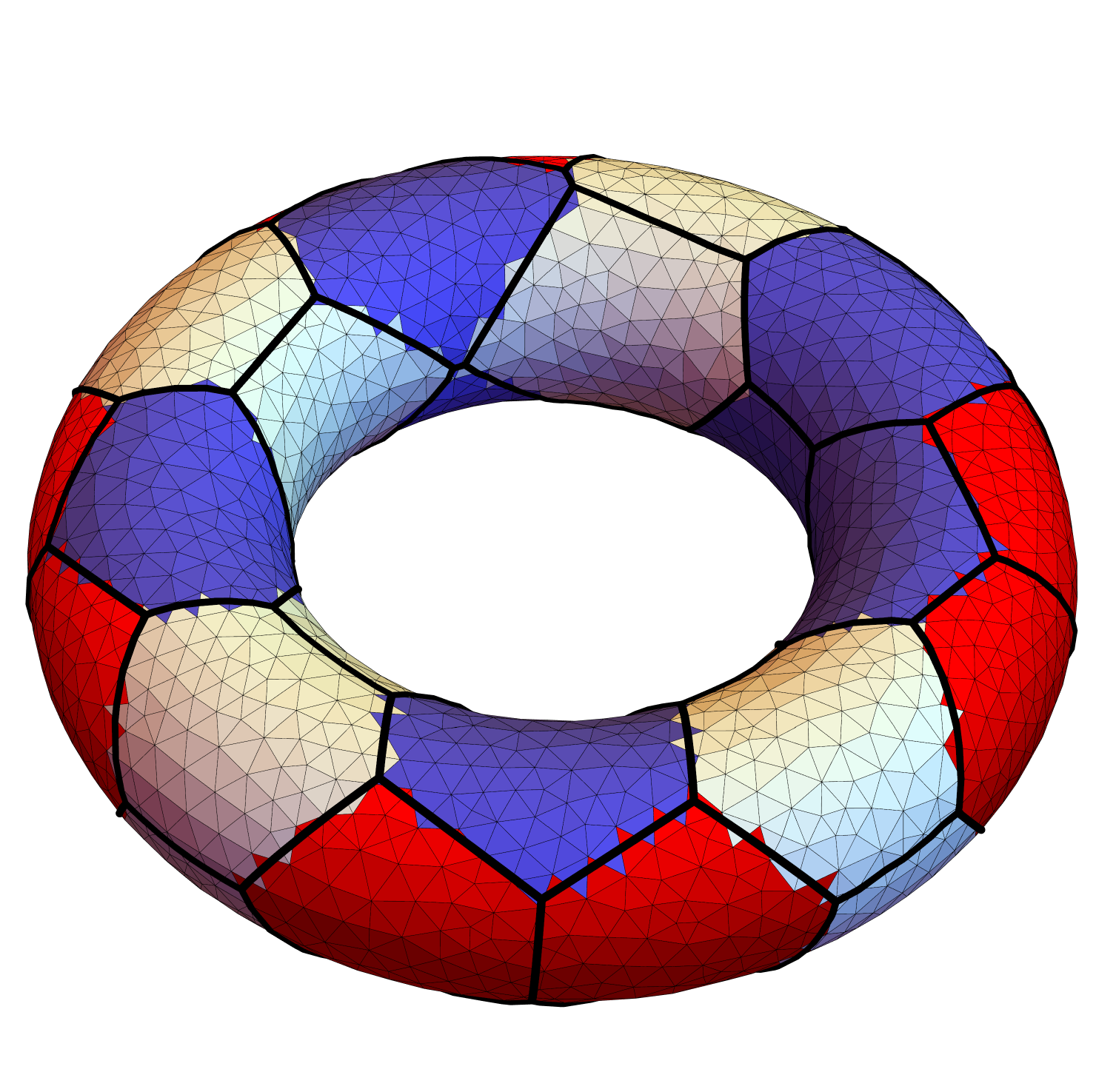}
\caption{The Voronoi diagram of a 32-particle local energy minima on the surface of the torus found via gradient descent. The boundaries of the Voronoi regions are shown with thick black lines, and the triangles of the underlying mesh are colored according to the connectivity of the particle nearest to their centroid in the surface Voronoi diagram: red regions correspond to particles with five neighbors, blue to particles with seven, and white to particles with six. The ground state found here has 10 five-seven defect pairs, whereas the ground state configuration predicted in Ref.~\cite{giomi2008elastic} has 16 five-seven defect pairs.}
\label{fig:smallTorusSmallNConfiguration}
\end{figure}

Ref.~\cite{giomi2008elastic} predicts, via a continuum elastic theory and direct particle-based numerical simulations using a long-range, repulsive force acting over Euclidean distance, a variety of interesting toroidal crystals -- these range from precise ground state configurations in the small-$N$ regime, to patterns of crystalline defects, to limiting torus aspect ratios beyond which defect-free configurations should be the true ground state of the system. In Fig.~\ref{fig:smallTorusSmallNConfiguration} we show an in-surface Voronoi diagram (generated as a restricted Euclidean Voronoi diagram via SurfaceVoronoi \cite{xin2022surfacevoronoi}) of a typical low-energy configuration. 

As gradient descent is particularly suitable for finding nearby \emph{local} minima rather than global minima,  in the tests below we primarily aim to find expected distributions of defects and qualitative geometric signatures, rather than true ground states. For example, given 32 particles on an aspect ratio 3 torus, the true ground state consists of 16 pairs of 5-7 defects -- 16 particles with 5 neighbors and 16 particles with 7 neighbors -- arranged on the outer and inner portions of the torus, respectively \cite{giomi2008elastic}. While gradient descent starting from a random point pattern does not find the true ground state, we indeed find a set of defect pairs distributed in the expected regions of the torus, as seen in  Fig.~\ref{fig:smallTorusSmallNConfiguration}.

\begin{figure}[b!]
\centering
\includegraphics[width=0.75\columnwidth]{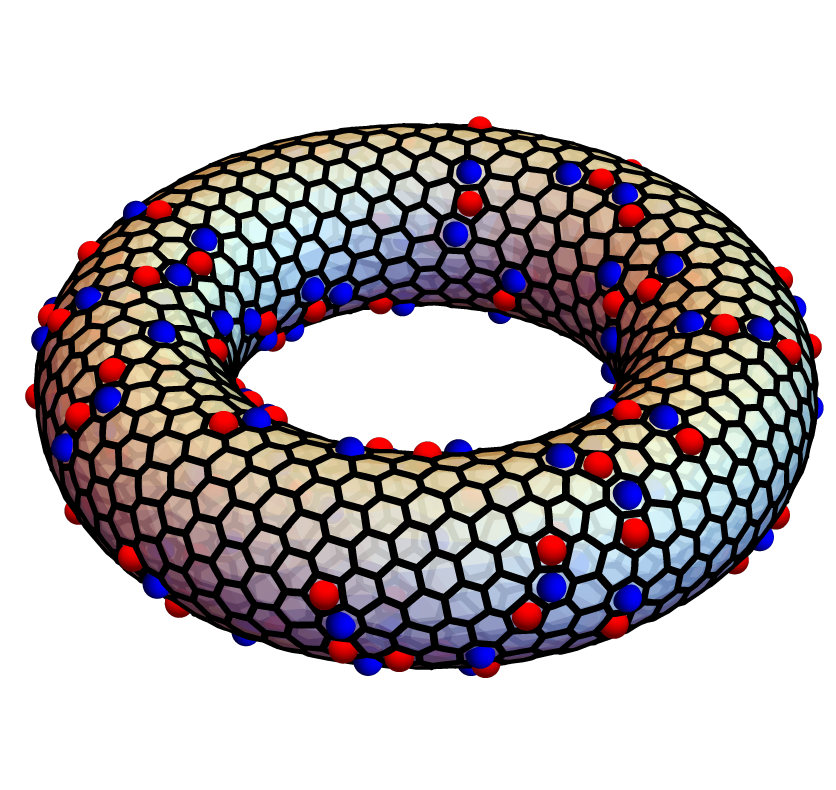}
\includegraphics[width=0.75\columnwidth]{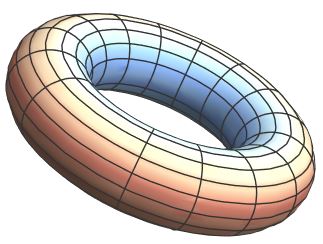}
\caption{ (Top) A relaxed configuration of 1000 particles on the surface of a torus. Here, rather than a strict localization of five-fold and seven-fold defects to the outer and inner equators of the torus as expected in the true ground state, we observe chains of defects populating the surface. (Bottom) A heat map indicating the average net defect charge across the surface (with red corresponding to an excess of five-fold defects and blue to an excess of seven-fold defects) over 1000 quenches of an $N=500$ particle configuration.}
\label{fig:smallTorusLargeN}
\end{figure}

As the number of particles on the surface of the torus increases, we expect to find \emph{chains} of defects coexisting with a large number of particles which interact with six neighbors \cite{giomi2008elastic}. We performed an $N=1000$ particle simulation which began from a random initial distribution of positions and relaxed via gradient descent. We observe a greater total number of defects than expected in the true ground state \cite{giomi2008elastic}, but as shown in the top panel of Fig.~\ref{fig:smallTorusLargeN} we indeed found the the qualitatively expected result that in this particle-number regime defects are primarily localized along chains. Leveraging the computational efficiency afforded by our simulation framework, we performed over 1000 separate energy minimizations of $N=500$ particles to find the expected distribution of defect charge in these local energy minima. As seen in the bottom panel of Fig.~\ref{fig:smallTorusLargeN}, we find that (as expected) there tends to be excesses of 7-fold and 5-fold defects on the inner and outer portions of the torus, and where the Gaussian curvature vanishes the expected defect charge also vanishes.

\begin{figure}[htb!]
\centering
\includegraphics[width=1\columnwidth]{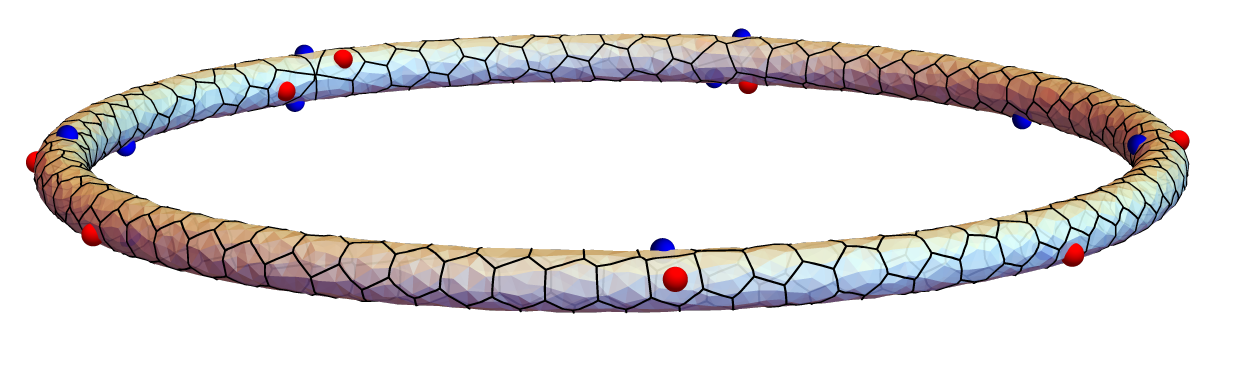}
\caption{A low-energy configuration of $N=300$ particles on a narrow (aspect ratio 20) torus. While gradient descent does not easily find the defectless ground state, this local minimum has relatively low defect number (11 five-fold and seven-fold defects, denoted with red and blue points, respectively).}
\label{fig:largeTorusConfigurationComparison}
\end{figure}

As the aspect ratio of the torus grows, the ground states correspond to lattices with a slow twist -- particles forming helical chains around the surface -- but with fewer total defects. We initially seed 300 particles positions on an aspect-ratio-20 torus on a lattice with \emph{incommensurate} twist (so that the initial configuration, while not disordered, is nevertheless far from the true ground state). We find that the particles relax to a structure with regions of correct crystalline order separated by grain boundaries. This is shown in Fig.~\ref{fig:largeTorusConfigurationComparison}. 
In detail, this configuration has 11 five-fold and seven-fold defects, many localized between different twisted lattice configurations (e.g., on the top left of the figure).

We have included supplemental videos showing the evolution of particle positions during minimization corresponding to the results shown in Figs.~\ref{fig:smallTorusSmallNConfiguration}-\ref{fig:largeTorusConfigurationComparison}. We have shown that we can reproduce qualitatively expected signatures of ground state configurations using direct gradient descent. This includes signatures of the high-defect-number ground state for 32 particles in their relaxed distributions on an aspect ratio 3 torus, defect chains and  expected distributions of defects over multiple quenches for many more particles on the same, and finally twisted regions within the aspect ratio 20 torus that reflect the expected helical configuration in the defectless ground state. Although not our goal in this section, we fully expect that implementing more sophisticated minimization protocols -- simulated annealing, quench-and-perturb, etc. -- would allow us to more fully reproduce the true ground-state behavior reported in the literature \cite{giomi2008elastic}.

\section{Discussion and future developments}\label{sec:conclusion}
We have a presented a new framework for simulating particles on curved surfaces. We build around existing exact geodesic solvers on discretized surfaces \cite{xin2009improving}, and implement new interfaces and data structures to efficiently submesh the surface when dealing with particles that interact via finite-range potentials. We integrate this with a standard, easily extensible object-oriented framework written in \CC{} to allow researches to easily adapt our code with specialized particle-particle interactions or dynamical update rules appropriate to their system. We have provided standard reports for the scaling of our implementation -- at fixed or varying particle number, on meshes of fixed or varying complexity, and when parallelizing across multiple processors -- so that other researchers may easily estimate whether simulations of a particular use-case may be achieved. 

We have chosen a particular, MMP-like geodesic distance solver, but we note that there are interesting differences in the computational complexity of alternate choices -- such as the vector heat method \cite{sharp2019vector} -- and the way exact vs approximate solvers lead to different discretization errors. We are currently exploring the implementation of other approaches, something our object-oriented design makes straightforward. 

Beyond this, natural extensions of our code base include the addition of vector-alignment interactions -- as  are common in models of flocking \cite{vicsek1995novel} -- to supplement the positional particle potentials already implemented. We have also currently focused on closed surfaces, but it is well known that some interesting phenomena can only manifest themselves on surfaces with boundaries \cite{giomi2007crystalline}. In this context and others it would be interesting to extend our \verb+shift+ algorithm to handle such boundary conditions, and to extend both it and the \verb+displacement+ algorithm to handle curved surfaces with periodic boundary conditions.

The most up-to-date version of our codebase, including branches dedicated to our graphical user interface and experimental future features, and a branch which allows the figure in this paper to be easily reproduced, can be found at \url{https://github.com/sussmanLab/curvedSpaceSim}.

\section{Acknowledgements}
We would like to thank Helen Ansell, Haicen Yue, and Tomilola Obadiya for fruitful discussions and critical comments on this manuscript. This material is based upon work supported by the National Science Foundation under Grant No.~DMR-2143815.

\bibliographystyle{elsarticle-num}
\bibliography{curvedSpaceSimulationsRefs}

\begin{thebibliography}{10}
\expandafter\ifx\csname url\endcsname\relax
  \def\url#1{\texttt{#1}}\fi
\expandafter\ifx\csname urlprefix\endcsname\relax\def\urlprefix{URL }\fi
\expandafter\ifx\csname href\endcsname\relax
  \def\href#1#2{#2} \def\path#1{#1}\fi

\bibitem{tarjus2011statistical}
G.~Tarjus, F.~Sausset, P.~Viot, Statistical mechanics of liquids and fluids in
  curved space, Advances in Chemical Physics 148 (2011) 251--310.

\bibitem{schamberger2023curvature}
B.~Schamberger, R.~Ziege, K.~Anselme, M.~Ben~Amar, M.~Bykowski, A.~P. Castro,
  A.~Cipitria, R.~A. Coles, R.~Dimova, M.~Eder, et~al., Curvature in biological
  systems: its quantification, emergence, and implications across the scales,
  advanced materials 35~(13) (2023) 2206110.

\bibitem{thomson1904xxiv}
J.~J. Thomson, Xxiv. on the structure of the atom: an investigation of the
  stability and periods of oscillation of a number of corpuscles arranged at
  equal intervals around the circumference of a circle; with application of the
  results to the theory of atomic structure, The London, Edinburgh, and Dublin
  Philosophical Magazine and Journal of Science 7~(39) (1904) 237--265.

\bibitem{tammes1930origin}
P.~M.~L. Tammes, On the origin of number and arrangement of the places of exit
  on the surface of pollen-grains, Recueil des travaux botaniques
  n{\'e}erlandais 27~(1) (1930) 1--84.

\bibitem{bowick2002crystalline}
M.~Bowick, A.~Cacciuto, D.~R. Nelson, A.~Travesset, Crystalline order on a
  sphere and the generalized thomson problem, Physical Review Letters 89~(18)
  (2002) 185502.

\bibitem{agarwal2020simple}
S.~Agarwal, S.~Hilgenfeldt, Simple, general criterion for onset of disclination
  disorder on curved surfaces, Physical review letters 125~(7) (2020) 078003.

\bibitem{agarwal2021predicting}
S.~Agarwal, S.~Hilgenfeldt, Predicting the characteristics of defect
  transitions on curved surfaces, Soft Matter 17~(15) (2021) 4059--4068.

\bibitem{bruss2012non}
I.~R. Bruss, G.~M. Grason, Non-euclidean geometry of twisted filament bundle
  packing, Proceedings of the National Academy of Sciences 109~(27) (2012)
  10781--10786.

\bibitem{bruss2013topological}
I.~R. Bruss, G.~M. Grason, Topological defects, surface geometry and cohesive
  energy of twisted filament bundles, Soft Matter 9~(34) (2013) 8327--8345.

\bibitem{pincus1984polymer}
P.~Pincus, C.~Sandroff, T.~Witten, Polymer adsorption on colloidal particles,
  Journal de Physique 45~(4) (1984) 725--729.

\bibitem{hanke1999critical}
A.~Hanke, S.~Dietrich, Critical adsorption on curved objects, Physical Review E
  59~(5) (1999) 5081.

\bibitem{liu2016curvature}
I.~B. Liu, N.~Sharifi-Mood, K.~J. Stebe, Curvature-driven assembly in soft
  matter, Philosophical Transactions of the Royal Society A: Mathematical,
  Physical and Engineering Sciences 374~(2072) (2016) 20150133.

\bibitem{goodwin2019smooth}
K.~Goodwin, S.~Mao, T.~Guyomar, E.~Miller, D.~C. Radisky, A.~Ko{\v{s}}mrlj,
  C.~M. Nelson, Smooth muscle differentiation shapes domain branches during
  mouse lung development, Development 146~(22) (2019) dev181172.

\bibitem{yu2021adaptive}
S.-M. Yu, B.~Li, F.~Amblard, S.~Granick, Y.-K. Cho, Adaptive architecture and
  mechanoresponse of epithelial cells on a torus, Biomaterials 265 (2021)
  120420.

\bibitem{chang2022quantifying}
Y.-W. Chang, R.~Cruz-Acu{\~n}a, M.~Tennenbaum, A.~A. Fragkopoulos, A.~J.
  Garc{\'\i}a, A.~Fern{\'a}ndez-Nieves, Quantifying epithelial cell
  proliferation on curved surfaces, Frontiers in Physics 10 (2022) 1055393.

\bibitem{marin2023mapping}
A.~Mar{\'\i}n-Llaurad{\'o}, S.~Kale, A.~Ouzeri, T.~Golde, R.~Sunyer,
  A.~Torres-S{\'a}nchez, E.~Latorre, M.~G{\'o}mez-Gonz{\'a}lez,
  P.~Roca-Cusachs, M.~Arroyo, et~al., Mapping mechanical stress in curved
  epithelia of designed size and shape, Nature communications 14~(1) (2023)
  4014.

\bibitem{luciano2024mechanoresponse}
M.~Luciano, M.~Versaevel, Y.~Kalukula, S.~Gabriele, Mechanoresponse of curved
  epithelial monolayers lining bowl-shaped 3d microwells, Advanced Healthcare
  Materials 13~(4) (2024) 2203377.

\bibitem{sausset2008tuning}
F.~Sausset, G.~Tarjus, P.~Viot, Tuning the fragility of a glass-forming liquid
  by curving space, Physical review letters 101~(15) (2008) 155701.

\bibitem{turci2017glass}
F.~Turci, G.~Tarjus, C.~P. Royall, From glass formation to icosahedral ordering
  by curving three-dimensional space, Physical Review Letters 118~(21) (2017)
  215501.

\bibitem{activesphere2015}
R.~Sknepnek, S.~Henkes, Active swarms on a sphere, Physical Review E
  91~(022306) (2015).

\bibitem{activepart2016}
Y.~Fily, A.~Baskaran, M.~F. Hagan, Active particles on curved surfaces,
  arXiv:1601.00324 [cond-mat.soft] (2016).

\bibitem{topsound2017}
S.~Shankar, M.~J. Bowick, M.~C. Marchetti, Topological sound and flocking on
  curved surfaces, Physical Review X 7~(031039) (2017).

\bibitem{brandstatter2023curvature}
T.~Brandst{\"a}tter, D.~B. Br{\"u}ckner, Y.~L. Han, R.~Alert, M.~Guo, C.~P.
  Broedersz, Curvature induces active velocity waves in rotating spherical
  tissues, Nature Communications 14~(1) (2023) 1643.

\bibitem{lavrentovich2016first}
M.~O. Lavrentovich, E.~M. Horsley, A.~Radja, A.~M. Sweeney, R.~D. Kamien,
  First-order patterning transitions on a sphere as a route to cell morphology,
  Proceedings of the National Academy of Sciences 113~(19) (2016) 5189--5194.

\bibitem{radja2019pollen}
A.~Radja, E.~M. Horsley, M.~O. Lavrentovich, A.~M. Sweeney, Pollen cell wall
  patterns form from modulated phases, Cell 176~(4) (2019) 856--868.

\bibitem{luciano2021cell}
M.~Luciano, S.-L. Xue, W.~H. De~Vos, L.~Redondo-Morata, M.~Surin, F.~Lafont,
  E.~Hannezo, S.~Gabriele, Cell monolayers sense curvature by exploiting active
  mechanics and nuclear mechanoadaptation, Nature Physics 17~(12) (2021)
  1382--1390.

\bibitem{pieuchot2018curvotaxis}
L.~Pieuchot, J.~Marteau, A.~Guignandon, T.~D. Santos, I.~Brigaud, P.~Chauvy,
  T.~Cloatre, A.~Ponche, T.~Petithory, P.~Rougerie, M.~Vassaux, J.~Milan, N.~T.
  Wakhloo, A.~Spangenberg, M.~Bigerelle, K.~Anselme, Curvotaxis directs cell
  migration through cell-scale curvature landscapes, Nature Communications
  9~(3995) (2018).

\bibitem{gehrels2023curvature}
E.~W. Gehrels, B.~Chakrabortty, M.-E. Perrin, M.~Merkel, T.~Lecuit, Curvature
  gradient drives polarized tissue flow in the drosophila embryo, Proceedings
  of the National Academy of Sciences 120~(6) (2023) e2214205120.

\bibitem{maechler2019curvature}
F.~A. Maechler, C.~Allier, A.~Roux, C.~Tomba, Curvature-dependent constraints
  drive remodeling of epithelia, Journal of cell science 132~(4) (2019)
  jcs222372.

\bibitem{tang2022collective}
W.~Tang, A.~Das, A.~F. Pegoraro, Y.~L. Han, J.~Huang, D.~A. Roberts, H.~Yang,
  J.~J. Fredberg, D.~N. Kotton, D.~Bi, et~al., Collective curvature sensing and
  fluidity in three-dimensional multicellular systems, Nature Physics 18~(11)
  (2022) 1371--1378.

\bibitem{happel2022effects}
L.~Happel, D.~Wenzel, A.~Voigt, Effects of curvature on epithelial
  tissue-coordinated rotational movement and other spatiotemporal arrangements,
  Europhysics Letters (2022).

\bibitem{rank2021active}
M.~Rank, A.~Voigt, Active flows on curved surfaces, Physics of Fluids
  33~(072110) (2021).

\bibitem{hueschen2023wildebeest}
C.~L. Hueschen, A.~R. Dunn, R.~Phillips,
  \href{https://link.aps.org/doi/10.1103/PhysRevE.108.024610}{Wildebeest herds
  on rolling hills: Flocking on arbitrary curved surfaces}, Phys. Rev. E 108
  (2023) 024610.
\newblock \href {https://doi.org/10.1103/PhysRevE.108.024610}
  {\path{doi:10.1103/PhysRevE.108.024610}}.
\newline\urlprefix\url{https://link.aps.org/doi/10.1103/PhysRevE.108.024610}

\bibitem{sussmancurvrigid2020}
D.~M. Sussman, Interplay of curvature and rigidity in shape-based models of
  confluent tissue, Physical Review Research 2~(023417) (2020).

\bibitem{thomas2023shape}
E.~C. Thomas, S.~Hopyan, Shape-driven confluent rigidity transition in curved
  biological tissues, Biophysical Journal 122~(21) (2023) 4264--4273.

\bibitem{giomi2008elastic}
L.~Giomi, M.~J. Bowick, Elastic theory of defects in toroidal crystals, The
  European Physical Journal E 27 (2008) 275--296.

\bibitem{hardin2005minimal}
D.~P. Hardin, E.~B. Saff, Minimal riesz energy point configurations for
  rectifiable d-dimensional manifolds, Advances in Mathematics 193~(1) (2005)
  174--204.

\bibitem{schonhofer2022curvature}
P.~W. Sch{\"o}nh{\"o}fer, S.~C. Glotzer, Curvature-controlled geometrical
  lensing behavior in self-propelled colloidal particle systems, Soft Matter
  18~(45) (2022) 8561--8571.

\bibitem{polyscope}
N.~Sharp, et~al., Polyscope, www.polyscope.run (2019).

\bibitem{frenkel2023understanding}
D.~Frenkel, B.~Smit, Understanding molecular simulation: from algorithms to
  applications, Elsevier, 2023.

\bibitem{schoenholz2021jax}
S.~S. Schoenholz, E.~D. Cubuk, Jax, md a framework for differentiable physics,
  Journal of Statistical Mechanics: Theory and Experiment 2021~(12) (2021)
  124016.

\bibitem{mishra2010fluctuations}
S.~Mishra, A.~Baskaran, M.~C. Marchetti, Fluctuations and pattern formation in
  self-propelled particles, Physical Review E 81~(6) (2010) 061916.

\bibitem{yamamoto2022non}
T.~Yamamoto, D.~M. Sussman, T.~Shibata, M.~L. Manning, Non-monotonic
  fluidization generated by fluctuating edge tensions in confluent tissues,
  Soft Matter 18~(11) (2022) 2168--2175.

\bibitem{muller1997simple}
F.~M{\"u}ller-Plathe, A simple nonequilibrium molecular dynamics method for
  calculating the thermal conductivity, The Journal of chemical physics
  106~(14) (1997) 6082--6085.

\bibitem{howard2016efficient}
M.~P. Howard, J.~A. Anderson, A.~Nikoubashman, S.~C. Glotzer, A.~Z.
  Panagiotopoulos, Efficient neighbor list calculation for molecular simulation
  of colloidal systems using graphics processing units, Computer Physics
  Communications 203 (2016) 45--52.

\bibitem{toukmaji1996ewald}
A.~Y. Toukmaji, J.~A. Board~Jr, Ewald summation techniques in perspective: a
  survey, Computer physics communications 95~(2-3) (1996) 73--92.

\bibitem{vicsek1995}
T.~Vicsek, A.~Czir\'ok, E.~Ben-Jacob, I.~Cohen, O.~Shochet,
  \href{https://link.aps.org/doi/10.1103/PhysRevLett.75.1226}{Novel type of
  phase transition in a system of self-driven particles}, Physical Review
  Letters 75 (1995) 1226--1229.
\newblock \href {https://doi.org/10.1103/PhysRevLett.75.1226}
  {\path{doi:10.1103/PhysRevLett.75.1226}}.
\newline\urlprefix\url{https://link.aps.org/doi/10.1103/PhysRevLett.75.1226}

\bibitem{gregoire2004onset}
G.~Gr{\'e}goire, H.~Chat{\'e}, Onset of collective and cohesive motion,
  Physical review letters 92~(2) (2004) 025702.

\bibitem{cavagna2015flocking}
A.~Cavagna, L.~Del~Castello, I.~Giardina, T.~Grigera, A.~Jelic, S.~Melillo,
  T.~Mora, L.~Parisi, E.~Silvestri, M.~Viale, et~al., Flocking and turning: a
  new model for self-organized collective motion, Journal of Statistical
  Physics 158 (2015) 601--627.

\bibitem{dombrowski2004self}
C.~Dombrowski, L.~Cisneros, S.~Chatkaew, R.~E. Goldstein, J.~O. Kessler,
  Self-concentration and large-scale coherence in bacterial dynamics, Physical
  review letters 93~(9) (2004) 098103.

\bibitem{sussman2017cellgpu}
D.~M. Sussman, cellgpu: Massively parallel simulations of dynamic vertex
  models, Computer Physics Communications 219 (2017) 400--406.

\bibitem{vuijk2021chemotaxis}
H.~D. Vuijk, H.~Merlitz, M.~Lang, A.~Sharma, J.-U. Sommer, Chemotaxis of
  cargo-carrying self-propelled particles, Physical Review Letters 126~(20)
  (2021) 208102.

\bibitem{post1986statistical}
A.~J. Post, E.~D. Glandt, Statistical thermodynamics of particles adsorbed onto
  a spherical surface. i. canonical ensemble, The Journal of chemical physics
  85~(12) (1986) 7349--7358.

\bibitem{kamien2002geometry}
R.~D. Kamien, The geometry of soft materials: a primer, Reviews of Modern
  physics 74~(4) (2002) 953.

\bibitem{needham2021visual}
T.~Needham, Visual differential geometry and forms: a mathematical drama in
  five acts, Princeton University Press, 2021.

\bibitem{jantzen2012geodesics}
R.~T. Jantzen, Geodesics on the torus and other surfaces of revolution
  clarified using undergraduate physics tricks with bonus: nonrelativistic and
  relativistic kepler problems, arXiv preprint arXiv:1212.6206 (2012).

\bibitem{crane2018discrete}
K.~Crane, Discrete differential geometry: An applied introduction, Notices of
  the AMS, Communication 1153 (2018).

\bibitem{crane2020survey}
K.~Crane, M.~Livesu, E.~Puppo, Y.~Qin, A survey of algorithms for geodesic
  paths and distances, arXiv preprint arXiv:2007.10430 (2020).

\bibitem{sharp2019vector}
N.~Sharp, Y.~Soliman, K.~Crane, The vector heat method, ACM Transactions on
  Graphics (TOG) 38~(3) (2019) 1--19.

\bibitem{mitchell1987discrete}
J.~S. Mitchell, D.~M. Mount, C.~H. Papadimitriou, The discrete geodesic
  problem, SIAM Journal on Computing 16~(4) (1987) 647--668.

\bibitem{chen1990shortest}
J.~Chen, Y.~Han, Shortest paths on a polyhedron, in: Proceedings of the sixth
  annual symposium on Computational geometry, 1990, pp. 360--369.

\bibitem{surazhsky2005fast}
V.~Surazhsky, T.~Surazhsky, D.~Kirsanov, S.~J. Gortler, H.~Hoppe, Fast exact
  and approximate geodesics on meshes, ACM transactions on graphics (TOG)
  24~(3) (2005) 553--560.

\bibitem{xin2009improving}
S.-Q. Xin, G.-J. Wang, Improving chen and han's algorithm on the discrete
  geodesic problem, ACM Transactions on Graphics (TOG) 28~(4) (2009) 1--8.

\bibitem{cgal:eb-23b}
{The CGAL Project}, \href{https://doc.cgal.org/5.6/Manual/packages.html}{{CGAL}
  User and Reference Manual}, {5.6} Edition, {CGAL Editorial Board}, 2023.
\newline\urlprefix\url{https://doc.cgal.org/5.6/Manual/packages.html}

\bibitem{polthier2006straightest}
K.~Polthier, M.~Schmies, Straightest geodesics on polyhedral surfaces, in: ACM
  SIGGRAPH 2006 Courses, Association for Computing Machinery, 2006, pp. 30--38.

\bibitem{gabriel2004open}
E.~Gabriel, G.~E. Fagg, G.~Bosilca, T.~Angskun, J.~J. Dongarra, J.~M. Squyres,
  V.~Sahay, P.~Kambadur, B.~Barrett, A.~Lumsdaine, et~al., Open mpi: Goals,
  concept, and design of a next generation mpi implementation, in: Recent
  Advances in Parallel Virtual Machine and Message Passing Interface: 11th
  European PVM/MPI Users’ Group Meeting Budapest, Hungary, September 19-22,
  2004. Proceedings 11, Springer, 2004, pp. 97--104.

\bibitem{cgal:klcdv-tsmsp-23b}
S.~Kiazyk, S.~Loriot, {\'E}.~C. de~Verdi{\`e}re,
  \href{https://doc.cgal.org/5.6/Manual/packages.html#PkgSurfaceMeshShortestPath}{Triangulated
  surface mesh shortest paths}, in: {CGAL} User and Reference Manual, {5.6}
  Edition, {CGAL Editorial Board}, 2023.
\newline\urlprefix\url{https://doc.cgal.org/5.6/Manual/packages.html#PkgSurfaceMeshShortestPath}

\bibitem{botsch2004remeshing}
M.~Botsch, L.~Kobbelt, A remeshing approach to multiresolution modeling, in:
  Proceedings of the 2004 Eurographics/ACM SIGGRAPH symposium on Geometry
  processing, 2004, pp. 185--192.

\bibitem{bowick2009two}
M.~J. Bowick, L.~Giomi, Two-dimensional matter: order, curvature and defects,
  Advances in Physics 58~(5) (2009) 449--563.

\bibitem{giomi2008defective}
L.~Giomi, M.~J. Bowick, Defective ground states of toroidal crystals, Physical
  Review E 78~(1) (2008) 010601.

\bibitem{xin2022surfacevoronoi}
S.~Xin, P.~Wang, R.~Xu, D.~Yan, S.~Chen, W.~Wang, C.~Zhang, C.~Tu,
  Surfacevoronoi: Efficiently computing voronoi diagrams over mesh surfaces
  with arbitrary distance solvers, ACM Transactions on Graphics (TOG) 41~(6)
  (2022) 1--12.

\bibitem{vicsek1995novel}
T.~Vicsek, A.~Czir{\'o}k, E.~Ben-Jacob, I.~Cohen, O.~Shochet, Novel type of
  phase transition in a system of self-driven particles, Physical review
  letters 75~(6) (1995) 1226.

\bibitem{giomi2007crystalline}
L.~Giomi, M.~Bowick, Crystalline order on riemannian manifolds with variable
  gaussian curvature and boundary, Physical Review B 76~(5) (2007) 054106.

\end{thebibliography}

\end{document}